\documentclass[pra,prl,twocolumn,showpacs,floatfix,10pt,twoside]{revtex4-1}
\usepackage{graphicx}
\usepackage{dcolumn}
\usepackage{bm}
\usepackage{subfigure}
\usepackage{float}
\usepackage{multirow}
\usepackage{booktabs}
\usepackage{amsmath}
\usepackage{latexsym,amssymb}
\usepackage[mathscr]{eucal}
\usepackage[switch*]{lineno}
\usepackage[bookmarks=true,
   colorlinks=true,
   linkcolor=blue,
   urlcolor=blue,
   citecolor=blue,
   bookmarks=true,
   hyperindex=true
]{hyperref}

\begin{document}
\allowdisplaybreaks[3]

\title{Modif\/ied fermion tunneling from higher-dimensional charged AdS black hole in massive gravity}

\author{Zhong-Wen Feng\textsuperscript{1,2}}
\altaffiliation{Email: zwfengphy@163.com}
\author{Qun-Chao Ding\textsuperscript{1,2}}
\author{Shu-Zheng Yang\textsuperscript{1,2}}

\vskip 0.5cm
\affiliation{1 Physics and Space Science College, China West Normal University, Nanchong, 637009, China\\
2 Department of Astronomy, China West Normal University, Nanchong 637009, China}


\begin{abstract}
The tunneling behavior of fermions with half-integral spin from a higher dimensional charged anti-de Sitter (AdS) black hole in de Rham, Gabadadze and Tolley (dRGT) massive gravity is investigated via a modif\/ied Hamilton-Jacobi equation. The results demonstrate that the modif\/ied thermodynamic
quantities not only are related  to the properties of the higher dimensional charged AdS black hole in dRGT massive gravity but also depend on the parameter $\beta$, the coupling constant $\sigma$ and the mass of emitted particles $m$. In addition, the modif\/ied Hawking temperature is higher than the original temperature; hence, the ef\/fect of MDR can signif\/icantly enhance the evolution of the black hole.
Besides,  our results can be verif\/ied using the modif\/ied  Stefan-Boltzmann law.
\end{abstract}
\keywords{Modif\/ied dispersion relation; Hawking radiation; Fermion tunneling; Hamilton-Jacobi equation}
\maketitle
\section{Introduction}
\label{Int}
Based on the classical viewpoint, black holes were once thought to only absorb objects \cite{ch0}. However, this changed after Hawking proved that black holes can radiate particles. In the theory of black hole radiation (which we now call it ``Hawking radiation theory''), Hawking introduced the quantum mechanism into gravity theory of curved spacetime and demonstrated that black holes can emit particles \cite{ch1,ch2}. This theory profoundly reveals the connection among quantum theory, gravitation theory, thermodynamics, and statistical physics. One can calculate the temperature of black holes using the Hawking radiation. Currently, the theory of Hawking radiation is considered the most important tool for investigating the thermodynamic properties of black holes.

Apart from the original research method of studying the radiation of black holes, many new methods have been proposed in the past forty years  \cite{ch3,ch3+,ch4,ch5,ch6,ch6+,ch6++}. In Refs. \cite{ch4,ch6}, Kraus, Parikh and Wilczek pointed out that the black hole radiation process can be considered as the quantum tunneling. Therefore, regarding the horizons as the tunneling barrier, one can easily obtain the tunneling rate of the emitted particles and the thermodynamic properties of black holes. Subsequently, Srinivasan and Padmanabhan developed the tunneling method and put forward the Hamilton-Jacobi ansatz, which substantially simplif\/ies the research process and promotes the understanding of black holes\cite{ch5}. Then, Kerner and Mann used the Hamilton-Jacobi ansatz to discuss fermion tunneling from spherically symmetric black holes \cite{ch6+,ch6++}. In Refs.~\cite{ch7+,ch7++}, Yang \emph{et al}. successfully derived the Hamilton-Jacobi equation from the Klein-Gordon equation, the Dirac equation, and the Maxwell equations. According to their works, the Hamilton-Jacobi ansatz can be used to describe the tunneling behavior of particles with spin on the horizons of black holes. Hence, the Hamilton-Jacobi ansatz is an ef\/fective approach for studying Hawking radiation. Using the Hamilton-Jacobi ansatz, extensive investigations of many types of black holes have been conducted \cite{ch7,ch8,ch9,ch9+,ch10,ch13+,ch14+,ch11,ch12,ch13}.

However, the classical theory of Hawking radiation has some defects \cite{ch14}. It has been found that the radiation that is derived from classical theory corresponds to a pure thermal state; hence, the black hole would emit all information at the end of evaporation, which would leads to the ``information loss paradox''. In addition, the singularities of spacetimes would be exposed to the universe since the horizons would disappear. For resolving this paradox, many methods have been proposed. In recent years, many works have claimed that there is a minimum measurable length in nature \cite{ch15,ch16,ch17}. In Refs.~\cite{ch18,ch19}, Amelino-Camelia showed that the standard energy-momentum dispersion relation must be changed to the modif\/ied dispersion relation (MDR)  nears the minimum measurable length. Based on the MDR, researchers have developed a new approach for overcoming the problems of the classical theory of Hawking radiation. In Refs.~\cite{ch20,ch21,ch22,ch23,ch23+,ch24+,ch25+,ch24,ch25}, the authors studied the modif\/ied thermodynamic properties of black holes via the MDR; according to their results, the MRD has signif\/icant ef\/fect on the evolution of black holes, namely, it prevents black holes from total evaporation and leads to remnants. Meanwhile, it is believed that the MDR can modify the equation of motion of particles on the event horizon of black holes. In Ref.~\cite{ch26,ch26a}, by using the MDR $p_0^2 = {\vec p^2} + {m^2} - {\left( {{\ell _p}{p_0}} \right)^{2 \beta}}{\vec p^2}$, which appears in space-time foam Liouville-string models, Kruglov obtained a  modif\/ied Dirac equation by f\/ixing $\beta=1$, that is,  $\left[ {{\gamma ^\mu }{\partial _\mu } + m - i{\ell _p}\left( {{{\bar \gamma }^t}{\partial _t}} \right)\left( {{{\bar \gamma }^i}{\partial _i}} \right)} \right]\psi  = 0$. Subsequently, when assessing $\beta=1$, Yang and his collaborators proposed another modif\/ied Dirac equation, namely, $\left[ {{\gamma ^\mu }{D_\mu } + {m \mathord{\left/ {\vphantom {m \hbar }} \right. \kern-\nulldelimiterspace} \hbar }} \right.$ $\left. { - \sigma \hbar \left( {{\gamma ^t}{D^t}} \right)\left( {{\gamma ^j}{D^j}} \right)} \right]\Psi  = 0$ and a new modif\/ied Klein-Gordon equation, that is, $\left( { - \partial _t^2 + \partial _j^2 + {m^2} - {\sigma ^2}{\hbar ^2}\partial _t^2\partial _j^2} \right) = 0$, where $\sigma$ is a very small parameter \cite{ch7+++,ch28}. Furthermore, all of those works claimed that the ef\/fect of MDR can be detected by  higher energy experiments.

Most of the works have been limited to the study of the tunneling behavior of particles with spin $0$ or $1/2$. However, it should be noted that the black holes radiate particles of both integral spin ($0$, $1$, $\cdots $ ) and half-integral spin (${1 \mathord{\left/ {\vphantom {1 2}} \right. \kern-\nulldelimiterspace} 2}$, ${3 \mathord{\left/ {\vphantom {3 2}} \right. \kern-\nulldelimiterspace} 2}$, $\cdots $ ). Therefore, in this paper, we discuss the modif\/ied tunneling behavior of fermions with half-integral spin. First, we derive the modif\/ied Hamilton-Jacobi equation from the Rarita-Schwinger equation in curved spacetime. Then, we study the fermion tunneling from a D-dimensional charged AdS black hole in dRGT massive gravity via the modif\/ied Hamilton-Jacobi equation. Finally, the correction for the Hawking temperature of the D-dimensional charged AdS black hole in dRGT massive gravity is obtained.

 The remainder of this paper is organized as follows: In Sec.~\ref{sec2}, according to a modif\/ied Dirac equation and the WKB approximation, we derive the modif\/ied Hamilton-Jacobi equation from the Rarita-Schwinger equation. In  Sec.~\ref{sec3}, by using the modif\/ied Hamilton-Jacobi equation, the modif\/ied tunneling rate of a fermion with half-integral spin and the correction for the Hawking temperature of a D-dimensional charged AdS black hole in dRGT massive gravity are obtained. The conclusions of this work are presented in  Sec.~\ref{Dis}.

\section{The modif\/ied Hamilton-Jacobi equation for fermions with half-integral spin}
\label{sec2}
In this section, we derive the modif\/ied Hamilton-Jacobi equation from the Rarita-Schwinger equation. In Ref.~ \cite{ch27}, the Rarita-Schwinger equation is as follows:
\begin{align}
\label{eq2}
\left( {{\bf{\bar \gamma }}^\mu  \partial _\mu   + \frac{m}{\hbar }} \right)\psi _{\alpha _1  \cdots \alpha _k }  = 0.
\end{align}
The above equation is used to describe the kinematics of fermions with half-integral spin in Minkowski spactime, which satisf\/ies the supplementary conditions  ${\bf{\bar \gamma }}^\mu  \psi _{\mu \alpha _{\rm{2}}  \cdots \alpha _k }  = \partial _\mu  \psi ^\mu  _{\alpha _{\rm{2}}  \cdots \alpha _\kappa }  = \psi ^\mu  _{\mu \alpha _{\rm{3}}  \cdots \alpha _\kappa }  = 0$, which satisfy the commutation relation ${\bf{\gamma }}^\mu  {\bf{\gamma }}^\upsilon   + {\bf{\gamma }}^\upsilon  {\bf{\gamma }}^\mu   = 2g^{\mu \upsilon } I$.  The role of the supplementary conditions is restrict the spin of fermions, for instance, if $\psi _{\alpha _{\rm{1}}  \cdots \alpha _{\rm{\kappa}} }  = \psi$, that is,  $\kappa=0$, the supplementary conditions vanish, and Eq.~(\ref{eq2}) becomes the Dirac equation, which describes the spin-$1/2$ fermion. However, if $\kappa=1$,  term $\psi ^\mu  _{\mu \alpha _{\rm{3}}  \cdots \alpha _\kappa }$ vanishes and Eq.~(\ref{eq2}) reduces to the kinematics equation for fermions with spin $3/2$ \cite{ch27}.

The spacetimes around the black holes are extremely curved. Therefore, in order to investigate the tunneling behavior of fermions on the event horizon of a black hole, one must generalize the Rarita-Schwinger equation to curved spacetime. In Refs.~\cite{ch7+++,ch28}, the Rarita-Schwinger equation in curved spacetime is given by
\begin{align}
\label{eq3}
\left( {{\bf{\gamma }}^\mu  \mathcal{D}_\mu   + \frac{m}{\hbar }} \right)\psi _{\alpha _1  \cdots \alpha _\kappa }  = 0.
\end{align}
Notably, the covariant derivative $\mathcal{D}_\mu$  in Eq.~(\ref{eq3}) is def\/ined as  $\mathcal{D}_\mu   \equiv \partial _\mu   + \Omega _\mu   + {{ieA_\mu  } \mathord{\left/ {\vphantom {{ieA_\mu  } \hbar }} \right. \kern-\nulldelimiterspace} \hbar }$ with the spin connection in curved spacetime  $\Omega _\mu$.

To obtain the modif\/ied corrected Hamilton-Jacobi equation for fermions with half-integral spin, we need to construct a deformed Rarita-Schwinger equation in curved spacetime by developing the viewpoints in Refs.~\cite{ch26,ch26a,ch7+++,ch28,ch27+}
\begin{align}
\label{eq5}
&\left[ {{{\bf{\gamma }}^\mu }{{\cal D}_\mu } + \frac{m}{\hbar } - \sigma {\hbar ^\beta }{{\left( {\sqrt {{g^{tt}}} {{\cal D}_t}} \right)}^{\beta  - 1}}\left( {{{\bf{\gamma }}^t}{{\cal D}_t}} \right)\left( {{{\bf{\gamma }}^j}{{\cal D}_j}} \right)} \right]
  \nonumber \\
& \cdot {\psi _{{\alpha _1} \cdots {\alpha _\kappa }}} = 0,
\end{align}
where parameter $\beta$ is a key characteristic of the magnitude of the ef\/fect of MDR, the value of which inf\/luences the modif\/ied results, and $\sigma$ is an extremely small coupling constant, which leads to the correction term $\sigma {\hbar ^\beta }{\left( {\sqrt {{g^{tt}}} {\mathcal{D}_t}} \right)^{\beta  - 1}}$ $\left( {{{\bf{\gamma }}^t}{\mathcal{D}_t}} \right) \left( {{{\bf{\gamma }}^j}{\mathcal{D}_j}} \right)$ being very small.  According to Eq.~(\ref{eq5}),  the wave function ${\psi _{{\alpha _1} \cdots {\alpha _\kappa }}}$ of a fermion with half-integral spin can be expressed as follows:
\begin{align}
\label{eq6}
{\psi _{{\alpha _1} \cdots {\alpha _\kappa }}} = {\xi _{{\alpha _1} \cdots {\alpha _\kappa }}}\exp \left( {{{iS} \mathord{\left/ {\vphantom {{iS} \hbar }} \right.
 \kern-\nulldelimiterspace} \hbar }} \right),
\end{align}
where $\xi _{\alpha _1  \cdots \alpha _\kappa }$ and $S$ are matrices and the action of the fermion, respectively \cite{ch20,ch27+,ch27+a,ch27a+,ch27a,ch27b,ch27c,ch27d,ch27d+,ch27d++,ch27d+++,ch27f,ch27g,ch27g+,ch27h,ch27h+}. The angular momentum parameters and the radiation energy parameters of radiation particles are denoted as $\partial _\varphi S =  j$ and $\partial _t S =  - \omega$, respectively. Now, substituting Eq.~(\ref{eq6}) into Eq.~(\ref{eq5}), and ignoring the higher order term  $\mathcal{O}\left( \hbar  \right)$, yields
\begin{align}
\label{eq7}
& \left\{ {i{\bf{\gamma }}^\mu  \left( {\partial _\mu  S + eA_\mu  } \right) + m - \sigma \left[ {\sqrt {g^{tt} } \left( { - i\omega  + ieA_t } \right)} \right]^{\beta  - 1} } \right.
  \nonumber \\
& \left. { \cdot {\bf{\gamma }}^t \left( {\omega  - eA_t } \right){\bf{\gamma }}^j \left( {\partial _j S + eA_j } \right)} \right\}\xi _{\alpha _1  \cdots \alpha _\kappa }  = 0.
\end{align}
By considering the relation  ${\bf{\gamma }}^\mu  \left( {\partial _\mu  S + eA_\mu  } \right) =  - {\bf{\gamma }}^t \left( {\omega  - eA_t } \right) + {\bf{\gamma }}^j \left( {\partial _j S + eA_j } \right)$, the equation above can be rewritten as
\begin{align}
\label{eq8}
&\Gamma ^\mu  \left( {\partial _\mu  S + eA_\mu  } \right)\xi _{\alpha _1  \cdots \alpha _\kappa  }
  \nonumber \\
&+ \left[ {m + \sigma \left( {g^{tt} } \right)^{\frac{{\beta  + 1}}{2}} \left( { - i\omega  + ieA_t } \right)^{\beta  + 1} } \right]\xi _{\alpha _1  \cdots \alpha _\kappa  }  = 0,
\end{align}
where $\Gamma ^\mu   = \gamma ^\mu   -\sigma \left( {g^{tt} } \right)^{\frac{{\beta  - 1}}{2}} \left( {-i\omega  + i eA_t } \right)^\beta  \gamma ^t \gamma ^\mu$. Multiplying $\Gamma ^\upsilon  \left( {\partial _\upsilon  S + eA_\upsilon  } \right)$  by Eq.~(\ref{eq8}), the results is
\begin{align}
\label{eq9}
&\Gamma ^\mu  \Gamma ^\upsilon  \Delta \xi _{\alpha _1  \cdots \alpha _\kappa  } + \left[ {m + \sigma \left( {g^{tt} } \right)^{\frac{{\beta  + 1}}{2}} \left( { - i\omega  + ieA_t } \right)^{\beta  + 1} } \right]^2
  \nonumber \\
&\cdot \xi _{\alpha _1  \cdots \alpha _\kappa  }  = 0,
\end{align}
where  $\Gamma^\mu  \Gamma^\upsilon    = \gamma ^\mu  \gamma ^\upsilon   -2 \sigma \left( {g^{tt} } \right)^{\frac{{\beta  - 1}}{2}} \left( {- i \omega  + i eA_t } \right)^\beta g^{t\mu } \gamma ^\upsilon   + {\cal O}\left( {\sigma ^2 } \right)$ and $\Delta  = \left( {\partial _\mu  S + eA_\mu  } \right)\left( {\partial _\upsilon  S + eA_\upsilon  } \right)$. Next, the subscripts  $\mu$ and $\upsilon$ are interchanged. Then, adding the result with Eq.~(\ref{eq9}), and multiplying it by $1/2$ \cite{ch28}, one yields
\begin{align}
\label{eq10}
& \left\{ {\frac{{\gamma ^\upsilon  \gamma ^\mu   + \gamma ^\mu  \gamma ^\upsilon  }}{2}\Delta  - \sigma \left( {g^{tt} } \right)^{\frac{{\beta  - 1}}{2}} } \right.\left( { - i\omega  + ieA_t } \right)^\beta
   \nonumber \\
&  \cdot  \left( {g^{\mu t} \gamma ^\upsilon   + g^{\upsilon t} \gamma ^\mu  } \right)\Delta + \left[ {m + \sigma \left( {g^{tt} } \right)^{\frac{{\beta  + 1}}{2}} \left( { - i\omega  + ieA_t } \right)^{\beta  + 1} } \right]^2
    \nonumber \\
& \left. {  + \mathcal{O}\left( {\sigma ^2 } \right)} \right\}\xi _{\alpha _1  \cdots \alpha _\kappa }  = \left\{ {g^{\mu \upsilon } \Delta  + m^2  + 2\sigma m\left( {g^{tt} } \right)^{\frac{{\beta  + 1}}{2}} } \right.
   \nonumber \\
& \cdot\left( { - i\omega  + ieA_t } \right)^{\beta  + 1}  - 2\sigma \left( {g^{tt} } \right)^{\frac{{\beta  - 1}}{2}} \left( { - i\omega  + ieA_t } \right)^\beta  g^{\upsilon t} \gamma ^\mu  \Delta
   \nonumber \\
& \left. { + \mathcal{O}\left( {\sigma ^2 } \right)} \right\}\xi _{\alpha _1  \cdots \alpha _\kappa  }  = 0.
\end{align}
Moreover, by using the anti-commutation relations of gamma matrices, Eq.~(\ref{eq10}) can be simplif\/ied as follows:
\begin{align}
\label{eq11}
i\sigma {\bf{\gamma }}^\mu  \left( {\partial _\mu  S + eA_\mu  } \right)\xi _{\alpha _1  \cdots \alpha _\kappa }  + \mathcal{M}  \xi _{\alpha _1  \cdots \alpha _\kappa }  = 0,
\end{align}
where
\begin{align}
\label{eq12}
{\cal M}   = \frac{{g^{\mu \upsilon } \Delta  + m^2  + 2\sigma m\left( {g^{tt} } \right)^{\frac{{\beta  + 1}}{2}} \left( {-i \omega  + i eA_t } \right)^{\beta  + 1} }}{{ - 2i\left( {g^{tt} } \right)^{\frac{{\beta  - 1}}{2}} \left( {- i\omega  + i eA_t } \right)^\beta  g^{\upsilon t} \left( {\partial _\upsilon  S + eA_\upsilon  } \right)}}.
\end{align}
Next, by multiplying the left side of Eq.~(\ref{eq11}) by $ - i\sigma {{\bf{\gamma }}^\upsilon }\left( {{\partial _\upsilon }S} \right.$ $\left. { + e{A_\upsilon }} \right)$, then, interchanging the subscripts $\mu$ and $\upsilon$, adding the result with Eq.~(\ref{eq11}) and dividing by $2$, the f\/inal result is obtained:
\begin{align}
\label{eq13}
&\left[ {\frac{{g^{\mu \upsilon } \Delta  + m^2  + 2\sigma m\left( {g^{tt} } \right)^{\frac{{\beta  + 1}}{2}} \left( { - i\omega  + ieA_t } \right)^{\beta  + 1} }}{{ - 2i\left( {g^{tt} } \right)^{\frac{{\beta  - 1}}{2}} \left( { - i\omega  + ieA_t } \right)^\beta  g^{\upsilon t} \left( {\partial _\upsilon  S + eA_\upsilon  } \right)}}} \right]^2
  \nonumber \\
& +\sigma ^2 g^{\mu \upsilon } \left( {\partial _\mu  S + eA_\mu  } \right)\left( {\partial _\upsilon  S + eA_\upsilon  } \right) = 0.
\end{align}
Here, the higher order terms, which are denoted as  $\mathcal{O}\left( {\sigma ^2 } \right)$ are ignored since they are very small, and one obtains
\begin{align}
\label{eq14}
&g^{\mu \upsilon } \left( {\partial _\mu  S + eA_\mu  } \right)\left( {\partial _\upsilon  S + eA_\upsilon  } \right)
 \nonumber \\
& + m^2  + 2\sigma m\left( {g^{tt} } \right)^{\frac{{\beta  + 1}}{2}} \left( {-i\omega  + i eA_t } \right)^{\beta  + 1}  = 0.
\end{align}
Obviously, Eq.~(\ref{eq14}) is the modif\/ied Hamilton-Jacobi equation for fermions with half-integral spin that was derived from the Rarita-Schwinger equation in curved spacetime. According to previous work, the Hamilton-Jacobi equation can be derived not only from the Rarita-Schwinger equation or the Dirac equation but also from the Klein-Gordon equation, the Maxwell equations, and the gravitational wave equation, among others. Therefore, the Hamilton-Jacobi equation can be used to describe the kinematic properties of particles with any spin. In the next section, we use modif\/ied Hamilton-Jacobi equation to investigate fermion tunneling from the D-dimensional charged AdS black hole in dRGT massive gravity.

\section{Fermions tunneling from the D-dimensional charged AdS black hole in dRGT massive gravity}
\label{sec3}
In this section, the tunneling behavior of fermions on the horizon of a D-dimensional charged AdS black hole in dRGT massive gravity is calculated. In higher dimensional dRGT massive gravity, the action with a negative cosmological constant and a Maxwell f\/ield can be expressed as
\begin{align}
\label{eq1+}
I = &\frac{1}{{16}}\int {{{\rm{d}}^D}x\sqrt { - g} }
\nonumber \\
&\cdot \left[ {\mathcal{R} - 2\Lambda  - \frac{1}{4}{ F_{\mu \nu } F^{\mu \nu }} + {m^2}\sum\limits_i^{n} {{c_i}{\mathcal{U}_i}\left( {g,f} \right)} } \right],
\end{align}
where $n\leq D-2$. The parameters $\mathcal{R}$, $c_i$ and $f$  are the scalar curvature,  a series of constants and a f\/ixed rank-2 symmetric tensor, respectively. $F_{\mu \nu }$ is the Maxwell f\/ield strength, which satisf\/ies  the relationship $F_{\mu \nu }={\partial_\mu} A_\nu -{\partial_\nu} A_\mu$. The last four terms of the above equation represent the massive potential  \cite{ch26+,ch29,ch30,ch30a,ch30b,ch31+,ch31a+}. Meanwhile,  when considering the higher dimensional massive gravity is ghost free, the symmetric polynomials of the eigenvalues ${{\cal U}_i}$ of the $D \times D$  matrix (${\cal K}_{{\kern 1pt} {\kern 1pt} {\kern 1pt} {\kern 1pt} {\kern 1pt} {\kern 1pt} \nu }^\mu  \equiv \sqrt {{g^{\mu \alpha }}{f_{\alpha \nu }}}$ )are \cite{ch30+,ch30++,ch30+++}
\begin{align}
\label{eq2+}
\mathcal{U}_1  = &\left[ \mathcal{K} \right],
\nonumber \\
\mathcal{U}_2  = &\left[ \mathcal{K} \right]^2  - \left[ {{\mathcal{K}^2 } } \right],
\nonumber \\
\mathcal{U}_3  = &\left[ \mathcal{K} \right]^3  - 3\left[ {\mathcal{K} } \right]\left[ {{\mathcal{K} }^2 } \right] + 2\left[ {{\mathcal{K} }^3 } \right],
\nonumber \\
 \mathcal{U}_4  = & \left[ \mathcal{K} \right]^4  - 6\left[ {{\mathcal{K}}^2 } \right]\left[ {\mathcal{K}} \right]^2  + 8\left[ {{\mathcal{K} }^3 } \right]\left[ {\mathcal{K} } \right]  + 3\left[ {\mathcal{K}^2 } \right]^2 - 6\left[ {{\mathcal{K} }^4 } \right],
\nonumber \\
 \mathcal{U}_5 = &\frac{1}{{60}}{\left[ {\cal K} \right]^5} - \frac{1}{6}{\left[ {\cal K} \right]^3}\left[ {{{\cal K}^2}} \right] + \frac{1}{3}{\left[ {\cal K} \right]^2}\left[ {{{\cal K}^3}} \right]- \frac{1}{3}\left[ {{{\cal K}^2}} \right]
 \nonumber \\
 & \cdot\left[ {{{\cal K}^3}} \right]  + \frac{1}{4}\left[ {\cal K} \right]{\left[ {{{\cal K}^2}} \right]^2} - \frac{1}{2}\left[ {\cal K} \right]\left[ {{{\cal K}^4}} \right] + \frac{2}{3}\left[ {{{\cal K}^5}} \right],
 \nonumber \\
\mathcal{U}_6 = &\frac{1}{{360}}{\left[ {\cal K} \right]^6} - \frac{1}{{24}}{\left[ {\cal K} \right]^4}\left[ {{{\cal K}^2}} \right] + \frac{1}{9}{\left[ {\cal K} \right]^3}\left[ {{{\cal K}^3}} \right] - \frac{1}{4}\left[ {{{\cal K}}} \right]^2\left[ {{{\cal K}^4}} \right]
 \nonumber \\
 &+ \frac{1}{8}\left[ {\cal K} \right]{\left[ {{{\cal K}^2}} \right]^2} - \frac{1}{{24}}{\left[ {{{\cal K}^2}} \right]^3} + \frac{1}{9}{\left[ {{{\cal K}^3}} \right]^2} - \frac{1}{3}\left[ {{{\cal K}^3}} \right]\left[ {{{\cal K}^2}} \right]
\nonumber \\
&\cdot \left[ {\cal K} \right] + \frac{1}{4}\left[ {{{\cal K}^4}} \right]\left[ {{{\cal K}^2}} \right] + \frac{2}{5}\left[ {{{\cal K}^5}} \right]\left[ {\cal K} \right] - \frac{1}{3}\left[ {{{\cal K}^6}} \right],
\nonumber \\
\mathcal{U}_7 = &\frac{1}{{2520}}{\left[ {\cal K} \right]^7} - \frac{1}{{240}}{\left[ {\cal K} \right]^5}\left[ {{{\cal K}^2}} \right] + \frac{1}{{72}}{\left[ {\cal K} \right]^4}\left[ {{{\cal K}^3}} \right]
       \nonumber \\
& - \frac{1}{{24}}{\left[ {\cal K} \right]^3}\left[ {{{\cal K}^4}} \right] + \frac{1}{{48}}{\left[ {\cal K} \right]^3}{\left[ {{{\cal K}^2}} \right]^2} - \frac{1}{{12}}{\left[ {\cal K} \right]^2}\left[ {{{\cal K}^2}} \right]
        \nonumber \\
& \cdot\left[ {{{\cal K}^3}} \right] + \frac{1}{{10}}{\left[ {\cal K} \right]^2}\left[ {{{\cal K}^5}} \right] - \frac{1}{{48}}{\left[ {{{\cal K}^2}} \right]^3}\left[ {\cal K} \right]+ \frac{1}{{24}}\left[ {{{\cal K}^2}} \right]
        \nonumber \\
&\cdot\left[ {{{\cal K}^5}} \right]  + \frac{1}{{18}}{\left[ {{{\cal K}^3}} \right]^2}\left[ {\cal K} \right] - \frac{1}{{12}}\left[ {{{\cal K}^3}} \right]\left[ {{{\cal K}^4}} \right]
\nonumber \\
& + \frac{1}{8}\left[ {{{\cal K}^6}} \right]\left[ {\cal K} \right] + \frac{1}{7}\left[ {{{\cal K}^7}} \right],
 \nonumber \\
\cdots \cdots &,
\end{align}
where $\mathcal{K}$  is the matrix square root, namely  $(\sqrt A )_{{\kern 1pt} {\kern 1pt} {\kern 1pt} \nu }^\mu (\sqrt A )_{{\kern 1pt} {\kern 1pt} {\kern 1pt} \lambda }^\nu  = A_{{\kern 1pt} {\kern 1pt} {\kern 1pt} \lambda }^\mu $, and the rectangular brackets denote the traces, namely,  $\left[ {\cal K} \right] = {\cal K}_{{\kern 1pt} {\kern 1pt} {\kern 1pt} \nu }^\mu $  and  $\left[ {{{\cal K}^n}} \right] = \left( {{K^n}} \right)_{{\kern 1pt} {\kern 1pt} {\kern 1pt} \nu }^\mu$  \cite{ch31}. Based on the action, which is expressed as Eq.~(\ref{eq1+}), the line element of a D-dimensional charged AdS black hole in dRGT massive gravity is expressed as
\begin{align}
\label{eq15}
{\rm d}s^2  =  - f\left( r \right){\rm d}t^2  + f\left( r \right)^{ - 1} {\rm d}r^2  + r^2 {\rm d}\Omega_{D-2} ^2,
\end{align}
where ${\rm d}\Omega_{D-2} ^2=h_{ij} {\rm d}x^i {\rm d}x^j$  represents the line element of a $\left( {D - 2} \right)$-dimensional space with constant curvature $\left( {D - 2} \right)\left( {D - 3} \right)k$, in which $k$ denotes the spatial curvature constant. When  $k=0$, the horizon hypersurface of spacetime is f\/lat. If the spatial curvature constant takes the value  $k=1$, the geometric property of black hole horizon hypersurface has positive curvature, whereas it has negative curvature if  $k=-1$. Considering reference metric $f_{\mu \nu }  = {\rm{diag}}\left( {0,0,c_0^2 h_{ij} } \right)$ and using the  notation ${d_k} = D - k$, the metric function $ f\left( r \right)$ can be expressed as follows:
\begin{align}
\label{eq16}
f\left( r \right) = & k + c_0^2{c_2}{m^2} + \frac{{{r^2}}}{{{l^2}}} + \frac{{{c_0}{c_1}{m^2}}}{{{d_2}}}r - \frac{{16\pi M}}{{{d_2}{V_{D - 2}}{r^{D - 3}}}}
\nonumber \\
& + \frac{{{d_3}c_0^3{c_3}{m^2}}}{r} + \frac{{{d_3}{d_4}c_0^4{c_4}{m^2}}}{{{r^2}}} + \frac{{{d_3}{d_4}{d_5}c_0^5{c_5}{m^2}}}{{{r^2}}}
 \nonumber \\
& + \frac{{{d_3}{d_4}{d_5}{d_6}c_0^6{c_6}{m^2}}}{{{r^2}}} + \frac{{{d_3}{d_4}{d_5}{d_6}{d_7}c_0^7{c_7}{m^2}}}{{{r^2}}}
 \nonumber \\
& + \frac{{{q^2}}}{{2{d_2}{d_3}{r^{2{d_3}}}}}+\cdots,
\end{align}
where $V_{D - 2}$  is the volume of space that is spanned by coordinates  $x_i$,  $M$  is the mass of the black hole \cite{ch29}, and $l$  represents the AdS radius. The electric f\/ield in a D-dimensional charged AdS black hole in dRGT massive gravity is ${F_{tr}} = {1 \mathord{\left/ {\vphantom {1 {{r^{{d_2}}}}}} \right. \kern-\nulldelimiterspace} {{r^{{d_2}}}}}$ and the electromagnetic potential becomes ${A_t} = {{ - q} \mathord{\left/ {\vphantom {{ - q} {\left( {{d_3}{r^{{d_3}}}{\rm{ }}} \right)}}} \right. \kern-\nulldelimiterspace} {\left( {{d_3}{r^{{d_3}}}{\rm{ }}} \right)}}$. Constant  $c_0$ satisf\/ies $c_0  > 0$ and terms $c_3m^2$  and $c_4m^2$  vanish if $D<5$  and  $D<6$, respectively. Term ${c_7}{m^2}$ only appears if $D\geq9$; hence, for convenience, we only consider $n=7$ and $D\geq 9$ in this paper \cite{ch30a+,ch30b+,ch30c+}. Based the geometric properties of the D-dimensional charged AdS black hole in dRGT massive gravity, its event horizon is located at  $\left. {f\left( r \right)} \right|_{r = r_H }  = 0$. Moreover, one can obtain the line element of the D-dimensional Schwarzschild AdS black hole solution when  $m\rightarrow 0$ and $q=0$. Based on the  ghost-free dRGT theory and  Eq.~(\ref{eq2+}), the form of $f\left( r \right)$ in the above equation includes additional graviton terms, which  rarely appear in previous works. Therefore, according to Eq.~(\ref{eq2+}), one may obtain more physical information about higher dimensional nonlinear massive gravity.

Now, substituting the inverse tensors of the  D-dimensional charged AdS black hole in dRGT massive gravity $g^{\mu \nu }$  into Eq.~(\ref{eq14}), the modif\/ied Hamilton-Jacobi equation can be rewritten as follows
\begin{align}
\label{eq17}
 - & f{\left( r \right)^{ - 1}}{\left[ {\frac{{\partial S}}{{\partial t}} - \frac{{eq}}{{\left( {D - 3} \right){r^{\left( {D - 3} \right)}}}}} \right]^2} + f\left( r \right){\left( {\frac{{\partial S}}{{\partial r}}} \right)^2} + {r^{ - 2}}
 \nonumber \\
 \cdot & {\left( {\frac{{\partial S}}{{\partial \Omega _{D - 2}^2}}} \right)^2} + 2\sigma mf{\left( r \right)^{\frac{{\beta  + 1}}{2}}}{\left[ { - i\omega  + i\frac{{eq}}{{\left( {D - 3} \right){r^{\left( {D - 3} \right)}}}}} \right]^{\beta  + 1}}
 \nonumber \\
 + & {m^2} = 0.
\end{align}
Considering the geometric properties of Eq.~(\ref{eq15}), the action can be represented as  $S =  - \omega t + W\left( r \right) + \Theta \left( \Omega_{D-2}  \right)$. Hence, the radial part of Eq.~(\ref{eq17}) is
\begin{align}
\label{eq18}
 - & f{\left( r \right)^{ - 1}}{\left[ {\omega  + \frac{{eq}}{{\left( {D - 3} \right){r^{\left( {D - 3} \right)}}}}} \right]^2} + f\left( r \right){\left[ {\frac{{\partial W\left( r \right)}}{{\partial r}}} \right]^2} + {m^2}
  \nonumber \\
 + & 2\sigma mf{\left( r \right)^{\frac{{\beta  + 1}}{2}}}{\left[ { - i\omega  + i\frac{{eq}}{{\left( {D - 3} \right){r^{\left( {D - 3} \right)}}}}} \right]^{\beta  + 1}} + {\lambda _0} = 0,
 \end{align}
and the nonradial part of the modif\/ied Hamilton-Jacobi equation becomes
\begin{align}
\label{eq19}
r^{ - 2} \left( {\frac{{\partial \Theta }}{{\partial \Omega _{D - 2}^2 }}} \right)^2  - \lambda _0  = 0,
\end{align}
where $\lambda _0$ is a separation constant. For a spherically symmetric black hole, the nonradial part of the modif\/ied Hamilton-Jacobi equation does not correspond to the tunneling rate of the emitted particles; hence, here, we only consider Eq.~(\ref{eq18}). It should be noted that the result of Eq.~(\ref{eq18}) depends on the value of $\beta$. Therefore, the MDR ef\/fect has an important inf\/luence on the tunneling behavior of fermions; similar conclusions can be found in Refs.~\cite{ch26,ch26a,ch7+++}. Hence, dif\/ferent values of $\beta$ lead to different tunneling rates. However, for convenience, we set $\beta=1$,  which leads to the integral function
\begin{align}
\label{eq20}
W_ \pm   = & \pm \int \frac{{dr}}{{f\left( r \right)}}
\nonumber \\
&\cdot \sqrt {{\left( {1 - 2\sigma m} \right)\left( {\omega  - eA_t } \right)^2  - f\left( r \right)\left( {m^2  + \lambda _0 } \right)} } ,
\end{align}
where $+(-)$ represent the outgoing (incoming) solutions, respectively. The solution to the above integral on the event horizon is
\begin{align}
\label{eq21}
W_ \pm   =  \pm i\pi \frac{{\left( {1 - m\sigma } \right)}}{{f'\left( {r_H } \right)}}\left( {\omega  - \omega _0 } \right),
\end{align}
where  $\omega _0  =  - {{(eq)} \mathord{\left/ {\vphantom {{eq} {\left[ {\left( {D - 3} \right)r_H^{D - 3} } \right]}}} \right. \kern-\nulldelimiterspace} {\left[ {\left( {D - 3} \right)r_H^{D - 3} } \right]}}$. Based on the tunneling theory of black holes, the tunneling rate of fermions with half-integral spin at the event horizon is given by
\begin{align}
\label{eq22}
\Gamma  &= \exp \left[ { - \frac{2}{\hbar }\left( {{\mathop{\rm Im}\nolimits} W_ +   - {\mathop{\rm Im}\nolimits} W_ -  } \right)} \right]
     \nonumber \\
&= \exp \left[ { - \frac{{4\pi }}{\hbar }\frac{{\left( {1 - m\sigma } \right ) \left( {\omega  - \omega _0 } \right)}}{{f'\left( {r_H } \right)}}} \right].
\end{align}
Since Eq.~(\ref{eq22}) is similar to the Boltzmann formula, the modif\/ied Hawking temperature of the D-dimensional charged AdS black hole in dRGT massive gravity becomes
\begin{align}
\label{eq23}
T_H  = \frac{\hbar}{{4\pi r_H }}\Xi \left( {1 + m\sigma } \right) = T_0 \left( {1 + m\sigma } \right),
 \end{align}
with
\begin{align}
\label{eq23+}
\Xi  =& {d_3}k + \frac{{{d_1}r_H^2}}{{{l^2}}} - \frac{{{q^2}}}{{2{d_3}r_H^{2{d_3}}}}  + {c_0}{c_1}{m^2}{r_H} + {d_3}{c_2}c_0^2{m^2}
\nonumber \\
& + \frac{{{d_3}{d_4}{c_3}c_0^3{m^2}}}{{{r_H}}} + \frac{{{d_3}{d_4}{d_5}{c_4}c_0^4{m^2}}}{{r_H^2}} + \frac{{{d_3}{d_4}{d_5}{d_6}{c_5}c_0^5{m^2}}}{{r_H^2}}
\nonumber \\
& + \frac{{{d_3}{d_4}{d_5}{d_6}{d_7}{c_6}c_0^6{m^2}}}{{r_H^2}} + \frac{{{d_3}{d_4}{d_5}{d_6}{d_7}{d_8}{c_7}c_0^7{m^2}}}{{r_H^2}} +\cdots.
 \end{align}
Besides, according to Ref.~\cite{ch31}, we express the mass of the D-dimensional charged AdS black hole in dRGT massive gravity in terms of the event horizon as follows
\begin{align}
\label{eq24}
M = &\frac{{{d_2}{V_{D - 2}}r_H^{{d_3}}}}{{16\pi }}\left[ {k + \frac{{r_H^2}}{{{l^2}}} + \frac{{{q^2}}}{{2{d_2}{d_3}r_H^{2{d_2}}}} + {c_0}{c_2}{m^2}} \right.
\nonumber \\
& + \frac{{{c_0}{c_1}{m^2}{r_H}}}{{{d_2}}} + \frac{{{d_3}c_0^3{c_3}{m^2}}}{{{r_H}}} + \frac{{{d_3}{d_4}c_0^4{c_4}{m^2}}}{{r_H^2}}
 \nonumber \\
& + \frac{{{d_3}{d_4}{d_5}c_0^5{c_5}{m^2}}}{{r_H^2}} + \frac{{{d_3}{d_4}{d_5}{d_6}c_0^6{c_6}{m^2}}}{{r_H^2}}
 \nonumber \\
& \left. { + \frac{{{d_3}{d_4}{d_5}{d_6}{d_7}c_0^7{c_7}{m^2}}}{{r_H^2}}} +\cdots \right].
 \end{align}
In Eq.~(\ref{eq23}),  $T_H$  is the modif\/ied Hawking temperature of the D-dimensional charged AdS black hole in dRGT massive gravity, whereas $T_0$  is the original temperature.

Furthermore, by using Eq.~(\ref{eq23}) and  the f\/irst law of black hole thermodynamics, namely, ${\rm{d}}S = \left( {{\rm{d}}M - {\Omega _H}{\rm{d}}J - } \right.$ ${{\left. {\Phi {\rm{d}}Q} \right)} \mathord{\left/ {\vphantom {{\left. {\Phi {\rm{d}}Q} \right)} T}} \right. \kern-\nulldelimiterspace} T}$ with the electromagnetic potential $\Omega_H$  and rotating potential $\Phi$, the modif\/ied entropy of the D-dimensional charged AdS black hole in dRGT massive gravity is expressed as
\begin{align}
\label{eq24+}
S_H & = \int {\frac{{{\rm d}M - \Phi {\rm d}Q}}{{\left( {1 + \sigma m} \right)T_0 }}}
     \nonumber \\
 & = S_0  - m\sigma \int {{\rm d}S_0 }  + \mathcal{O}\left( {\sigma ^2 } \right),
 \end{align}
where ${{S}}_0  = {{V_{D - 2} r_H^{D - 2} } \mathord{\left/ {\vphantom {{V_{D - 2} r_H^{D - 2} } 4}} \right. \kern-\nulldelimiterspace} 4}$ is the original entropy of the D-dimensional charged AdS black hole in dRGT massive gravity. It is clear that both modif\/ied Hawking temperature $T_H$ and entropy $S_H$ not only are related  to the properties of the D-dimensional charged AdS black hole in dRGT massive gravity but also depend on the coupling constant $\sigma$  and the mass of the emitted particles  $m$. Our results more accurately describe the thermodynamic properties of  charged AdS spacetime in dRGT gravity since the graviton terms are included. However, if $\sigma=0$, those modif\/ications reduce to $T_0$ and  $S_0$.

According to Eq.~(\ref{eq24}), the modif\/ied Hawking temperature is higher than the original temperature. Therefore, the ef\/fect of MDR can substantially enhance the evolution of the black hole. Hence, according to our result, MDR can accelerate the emission of particles from black holes. The emission rate of black holes is described by the Stefan-Boltzmann law. Therefore, it is interesting to study the ef\/fect of MDR on the Stefan-Boltzmann law, which is expressed as follows:
\begin{align}
\label{eq25}
\frac{{dE}}{{dt}} = \varrho _S AT^4,
\end{align}
where $E$  is the total energy of the black hole, $\varrho_S$  is the Stefan-Boltzmann constant,  and $A$  and  $T$ represent the area and temperature of the black hole, respectively. Now, by inserting Eq.~(\ref{eq24}) into Eq.~(\ref{eq25}), the Stefan-Boltzmann law can be rewritten as
\begin{align}
\label{eq26}
\left( {\frac{{dE}}{{dt}}} \right)_{{\rm{modif\/ied}}} & = \sigma _S AT_0^4 \left( {1 + m\sigma } \right)^4
     \nonumber \\
&= \left( {\frac{{dE}}{{dt}}} \right)_{{\rm{original}}} \left( {1 + m\sigma } \right)^4,
\end{align}
where  $\left( {{{dE} \mathord{\left/ {\vphantom {{dE} {dt}}} \right. \kern-\nulldelimiterspace} {dt}}} \right)_{{\rm{original}}}$ is the original Stefan-Boltzmann law. According to Eq.~(\ref{eq26}), the modif\/ied emission rate is higher than the original rate, which accords with a previous analysis of the Hawking temperature. The above results are obtained under the assumption $\beta=1$; if once $\beta$ takes another value, the f\/inal result will change accordingly.

\section{Discussion}
\label{Dis}
In this paper, according to the MDR ef\/fect and the Rarita-Schwinger equation, we constructed a  deformed Hamilton-Jacobi equation in curved spacetime, which can be used to describe the kinetic characteristics of fermions with half-integral spin.  Subsequently, the tunneling behavior of fermions with half-integral spin from a higher-dimensional AdS black hole in dRGT massive gravity is investigated via this modif\/ied Hamilton-Jacobi equation. It is found that if we f\/ix $\beta=1$, the modif\/ied thermodynamic quantities not only are related to the properties of spacetime of the higher dimensional charged AdS black hole in dRGT massive gravity but also depend on the coupling constant $\sigma$  and the mass of the emitted particles $m$.  Meanwhile, we calculated the modif\/ied emission rate by using Stefan-Boltzmann's law. Our results demonstrate that the modif\/ied values are higher than the original values; hence, the ef\/fect of MDR can substantially enhance the evolution of the black hole. Notably, based on ghost-free dRGT theory, the graviton terms, namely, ${{\cal U}_i}$ can be extended to  higher dimensions, which leads to the form of $f\left( r \right)$ in Eq.~(\ref{eq16}), which dif\/fers from those in previous works. Therefore, one may obtain more physical information about higher dimensional nonlinear massive gravity via our results. In addition, this work demonstrates that the ef\/fect of MDR has a substantial inf\/luence on the deformed Hamilton-Jacobi equation: the modif\/ied tunneling rate of fermions, the Hawking temperature, the entropy and the Stefan-Boltzmann law depend on the value of parameter $\beta$.  It is believed that interesting and meaningful results can be obtained from this f\/inding, which we will discuss in detail in future works.

\section*{Acknowledgements}
The authors thank Prof. Kai Lin for helpful suggestions and enlightening comments, which helped to improve the quality of this paper.

\end{document}